\documentclass[format=sigconf, review=false, anonymous=false]{acmart}

\AtBeginDocument{%
  \providecommand\BibTeX{{%
    \normalfont B\kern-0.5em{\scshape i\kern-0.25em b}\kern-0.8em\TeX}}}

\setcopyright{acmcopyright}

\acmConference[FAccT '21]{Conference on Fairness, Accountability, and Transparency}{XX}{XX}

\usepackage{lipsum}
\usepackage{subcaption}
\usepackage{array}
\newcommand{\PreserveBackslash}[1]{\let\temp=\\#1\let\\=\temp}
\newcolumntype{C}[1]{>{\PreserveBackslash\centering}p{#1}}
\newcolumntype{R}[1]{>{\PreserveBackslash\raggedleft}p{#1}}
\newcolumntype{L}[1]{>{\PreserveBackslash\raggedright}p{#1}}

\usepackage{stfloats}

\usepackage{color-edits}
\addauthor[Alex]{ac}{purple}
\addauthor[Nil-Jana]{na}{blue}

\setcopyright{rightsretained} 
\begin{document}
\copyrightyear{2021}
\acmYear{2021}
\acmConference[FAccT '21]{Conference on Fairness, Accountability, and Transparency}{March 3--10, 2021}{Virtual Event, Canada}
\acmBooktitle{Conference on Fairness, Accountability, and Transparency (FAccT '21), March 3--10, 2021, Virtual Event, Canada}\acmDOI{10.1145/3442188.3445877}
\acmISBN{978-1-4503-8309-7/21/03}

\title{The effect of differential victim crime reporting on predictive policing systems}

\author{Nil-Jana Akpinar}
\affiliation{%
    nakpinar@stat.cmu.edu\\
    Department of Statistics and Data Science \& Machine Learning Department\\
    Carnegie Mellon University
}

\author{Maria De-Arteaga}
\affiliation{
Information, Risk, and Operations Management Department\\
    McCombs School of Business\\
    University of Texas at Austin
}

\author{Alexandra Chouldechova}
\affiliation{
    Heinz College \& Department of Statistics and Data Science\\
    Carnegie Mellon University
    }


\begin{abstract}
    Police departments around the world have been experimenting with forms of place-based data-driven proactive policing for over two decades.  Modern incarnations of such systems are commonly known as hot spot predictive policing.
    These systems predict where future crime is likely to concentrate such that police can allocate patrols to these areas and deter crime before it occurs.
    Previous research on fairness in predictive policing has concentrated on the feedback loops which occur when models are trained on discovered crime data, but has limited implications for models trained on victim crime reporting data.
    We demonstrate how differential victim crime reporting rates across geographical areas can lead to outcome disparities in common crime hot spot prediction models.
    Our analysis is based on a simulation\footnote{Code available at \url{https://github.com/nakpinar/diff-crime-reporting.git}.} patterned after district-level victimization and crime reporting survey data for Bogot\'a, Colombia.
    Our results suggest that differential crime reporting rates can lead to a displacement of predicted hotspots from high crime but low reporting areas to high or medium crime and high reporting areas. This may lead to misallocations both in the form of over-policing and under-policing.
\end{abstract}

\maketitle

\section{Introduction}


Police departments around the world have been experimenting with computer-aided place-based predictive policing systems for over two decades.  In a 1998 National Institute of Justice survey, 36\% of police agencies employing over 100 sworn officers reported having the computing capability and data infrastructure to digitally generate crime maps \cite{mamalian1999use}.  Just a few years later, over 70\% of  agencies reported using such maps to identify crime hot spots as part of a broader adoption of CompStat approaches to policing \cite{weisburd2008compstat}.   More modern incarnations of predictive policing date back to 2008, when the  Los Angeles Police Department (LAPD) began its explorations of these systems, followed shortly thereafter by efforts such as the New York Police Department's use of tools developed by firms including Azavea, KeyStats and PredPol (2012+).  Far from being a US-centric phenomenon, such systems are widely used throughout Europe, the UK, and China \cite{jansen2018data,Babuta2020,Sprick2020}.

More recently, predictive policing systems have come under scrutiny due to their lack of transparency \cite{winston_2018} and concerns that they may lead to further over-policing of minority communities by virtue of being trained on biased or ``dirty'' data \cite{Lum2016, Ensign2018, Richardson2019}.  Critics commonly point to the possibility that such systems may produce dangerous feedback loops, vicious cycles wherein data on recent arrests is used to deploy police in still greater numbers to neighbourhoods where they zealously seek out suspicious activity and conduct even more arrests.  Recent work by \citet{Lum2016} and \citet{Ensign2018} has demonstrated both empirically and theoretically how such feedback loops can arise. 

Proponents and developers of predictive policing technologies have argued that such analyses are based on models of crime and policing that do not accurately reflect the types of data used as inputs to such systems, nor the types of crime that they seek to predict.  The analysis of \citet{Lum2016}, for instance, convincingly demonstrates how using data on drug arrests in Oakland, CA as inputs to a self-exciting point process (SEPP) model of the kind used in PredPol would result in high concentrations of policing in racial and ethnic minority neighbourhoods.  Yet PredPol has stated that they do not use data on drug-related offenses (or traffic citations) in generating their predictions, nor do they use data on arrests \cite{predpol_2017}.  Azavea, the creators and former owners of the HunchLab product, likewise note that their models focus on property and violent crimes, and the crime data they use is based on victim reporting rather than arrests \cite{cheetham_2019}.  

Secondly, proponents and developers have argued that prior studies incorrectly assume that targeted policing strategies lead to an escalation in crime detection and, correspondingly, arrests.  However, the adoption of hot spot policing strategies is predicated on an anticipated \textit{deterrence} effect.  Studies of the impacts of predictive policing on property and violent crimes and on arrests at targeted locations have produced mixed results.  A 2014 analysis of a randomized controlled experiment (RCT) conducted by RAND in Shreveport, Louisiana found no statistical evidence of crime reduction in the prediction-targeted locations compared to control locations \cite{Hunt2014}.  Another RCT conducted in Pittsburgh reported a 34\% drop in serious violent crime in ``temporary hot spots'' and a 24\% drop in ``chronic hot spots'' \cite{Fitzpatrick2018}.  This study found no evidence of crime displacement to nearby locations, and reported that a total of 4 arrests took place during the experiment's 20,000 hot spot patrols.  A peer-reviewed study published by researchers affiliated with PredPol concluded that, while arrests were higher at predicted locations, they were lower or comparable once the counts were adjusted for differences in crime rate \cite{brantingham2018does}.  PredPol has reported crime drops ranging from 8-30\% depending on the jurisdiction and type of crime \cite{predpol_results_2017}. 
While none of these counterarguments establish (or even claim) that the victim crime reporting data used to inform predictive policing systems is free from bias or leads to unbiased practices, they do point to a need for further investigation in settings that more closely mirror standard practice.   Our work presents an initial step in this direction.

In this paper we empirically demonstrate how predictive policing systems trained exclusively on victim crime reporting data (rather than arrest data) may nevertheless suffer from significant biases due to variation in reporting rates.
Our analysis is based on a simplified crime simulation patterned after district-level crime statistics for Bogot\'a, Colombia released by Bogot\'a's chamber of commerce, C\'amara de Comercio de Bogot\'a (CCB).
We demonstrate that variation in crime reporting rates can lead to significant mis-allocation of police.  These findings  corroborate the effects of differential victim crime reporting on predictive policing models hypothesized in \cite{maria_2020_crime}. We also discuss the limitations of using reporting rates from existing crime victimization surveys to attempt to correct for such biases.

\section{Background \& Related Work}

\subsection{Feedback loops and other biases in predictive policing}

Having already described the work of \citet{Lum2016}, we focus here on \cite{Ensign2018}. \citet{Ensign2018} theoretically characterize why feedback loops occur by modeling arrest-based predictive policing systems via a generalized P\'olya urn model.  Their analysis also considers a scenario in which both reported and detected crimes (i.e., arrests) are used to update beliefs about existing crime rates.  In the latter case they show that if the reported crime rates are an accurate reflection of underlying crime, then feedback loops can be avoided if either (a) underlying crime rates across regions are uniform to begin with; or (b) detected crimes aren't considered at all.  As we will discuss, there is considerable variation in the extent to which reported crimes reflect true underlying crime levels.    

\citet{Richardson2019} present three case studies where there is evidence that ``dirty data'' may have biased the targets of predictive policing systems.  Their case studies focus primarily on person-based predictive policing systems.  In the case of Maricopa County, Arizona, however, the authors report one instance in which biased data may have informed a PredPol system used by the Mesa Police Department and an RTMDx system used by the Glendale Police Department.  As the authors note, due to the lack of transparency surrounding what data was used and how, it is difficult to draw definitive conclusions.  This, however, does not make the documented patterns of biased practices against Maricopa County's Latino residents any less concerning.  

\subsection{Victim crime reporting}

Many countries and local governments conduct crime victimization surveys to better understand factors that drive differences in crime reporting rates, and to assess discrepancies between official crime statistics and victimization-based measures of criminal activity.  According to the 2018 report released by the Bureau of Justice Statistics, which oversees the annual US National Crime Victimization Survey (NCVS), 61\% of aggravated assaults, 63\% of robberies, 38\% of simple assaults, and only 25\% rapes/sexual assaults are reported to police \cite{morgan2018criminal}.  In this section we briefly overview different sources of disparities in victim crime reporting in the US context.  We note that, while our data simulation is based on a 2014 survey conducted in Bogot\'a---and crime reporting rates are observed to be considerably lower there---a number of our conclusions apply to geography-associated disparities in reporting rates in general.  In particular, our analysis indicates that, to the extent that these sources of disparity coincide with geography, we can expect significant under- or over-targeting to result.

The likelihood that a crime is reported to police has been found to be greater for older victims~\cite{hashima1999violent, watkins2005examining, bosick2012reporting, baumer2002neighborhood} and when the victim is a woman~\cite{baumer2010reporting}.  
It is also greater if a third party is present ~\cite{baumer2010reporting}, if a weapon is present or the victim is injured~\cite{baumer2010reporting, xie2006prior}.  Furthermore, reporting rates tend to increase with the degree to which the victim is of higher socioeconomic status than the offender, which in part accounts for the greater likelihood of white victims reporting crimes perpetrated by black offenders for crimes such as assaults~\cite{xie2012racial}.  However, \citet{xie2012racial} also observe that black-on-black assaults had by far the highest reporting rate in their study (44\%, compared to 25-33\% for other racial pairs).  This finding of high reporting rates for intra-racial black-on-black crimes was also reported in \cite{avakame1999did}.  In other words, while some might expect reporting rates to be lowest in predominantly black communities, this does not appear to be borne out by the data.
Furthermore, the degree of neighborhood socioeconomic disadvantage is not consistently associated with the likelihood of crime reporting~\cite{baumer2002neighborhood}.  An association has been observed for simple assaults, but not for robbery or aggravated assault. 
An extensive review of research in victim crime reporting is given in \cite{Xie2019review}.

There are many reasons for why particular incidents may not be reported to police.  These include fear of repercussion, victim perceptions that their victimization was `trivial', or might be perceived as such by police, or personal relationships with the offender.  Furthermore, there are documented examples of police actively discouraging victims from filing complaints in order to deflate serious crime statistics  \cite{Richardson2019}.    

\subsection{Predictive policing models}

The literature on predictive policing has considered a range of different modeling approaches for spatio-temporal crime forecasting and hot spot selection \cite{Fitzpatrick2019}.
To the best of our knowledge, only a small subset of models have been deployed and evaluated in practice.

PredPol, one of the largest vendors of predictive policing software in the US, has been one of few companies to publish modeling details of their hot spot prediction algorithm \cite{Mohler2011,Mohler2014,Mohler2015}.
The PredPol algorithm relies on a Self-Exciting Spatio-Temporal Point Process (SEPP) model that uses the location and time of historical incidents to predict the spatio-temporal distribution of future crime within a city.
Hot spot predictions for subsequent time steps can be obtained by evaluating the predicted crime distribution on a grid of cells overlaying the city.
This model, which has its roots in seismology, separates crime occurrences into ``background crime'' and ``offspring crime'' with the rationale that, similar to earthquakes which often trigger close-by aftershock earthquakes, crime tends to form clusters in time and space with burglars returning to the same areas or gang conflicts leading to retaliatory violence \cite{Mohler2011}.

While the SEPP method models both the space and time distribution explicitly, many other common approaches focus of one component at a time.  For instance, one straightforward approach is to apply time-series analysis to forecast crime counts in small pre-defined  spatial units such as individual segments of streets or grid cells.
In a field experiment with the Pittsburgh Bureau of Police, \citet{Fitzpatrick2018} used a within-cell moving average of crime counts in order to predict chronic hot spots and a within-cell multi-layer perceptron on lagged crime count features to predict temporary crime flare-ups. 
The authors report that the relatively simple moving average model alone was able to capture more crime on average than other models including SEPP models.
The Shreveport Police Department in Louisiana conducted experiments with a logistic regression model in 2012 \cite{Hunt2014}.
In addition to different lagged crime counts, predictors also included the number of juvenile arrests in the past six month and the presence of residents on probation and parole in each of the 400-by-400-foot grid cells.
Other methods focus on the spatial distribution of crime and aggregate the temporal component.
Spatial kernel density estimates (KDE's) and risk terrain modelling, which involves risk factors beyond crime rates, are used to help identify chronic hot spots but generally require visual inspection if spatial discretization is to be avoided \cite{Gorr2014, Fitzpatrick2019}.  A number of these models including SEPP's \cite{dulce2018efficient}, KDE's, moving-average type models, and other approaches  \cite{barreras2016comparison, dulce2018predicting} have previously been applied to historical crime data from Bogot\'a.  \citet{barreras2016comparison} found, for instance, that KDE models performed the best in their analysis.  

In this study, we focus on SEPP models for crime hot spot prediction as they appear to be one of the most widely used and analysed type of model, a trend driven in part by PredPol's popularity and the descriptions of their models publicly available in peer-reviewed literature.
For comparison purposes, we also consider a moving average model as analysed in \cite{Fitzpatrick2018}.
Both models are based only on the location an time of previous crimes, which makes them particularly accessible to police departments.

\section{Methodology}
\subsection{Self-Exiting Spatio-Temporal Point Processes}

Self-Exciting Spatio-Temporal Point Processes (SEPP) are a commonly used class of models for applications in which the rate of events depends on nearby past events, e.g. modeling of earth quakes or the spread of infectious diseases. 
In the purely temporal case, this class of models is also known as Hawkes processes.
We give a short introduction to SEPP, the specifications used in predictive policing and the model used in this study.
A more detailed review of SEPP can be found in \cite{Reinhart2018}.

SEPPs separate events into two types: background events and offspring events. 
Background events are generally assumed to occur independently across space and time according to a Poisson point process.
Each event can then cause offspring events in its vicinity according to a triggering function decaying in space and time.
The rate of events at locations $(x,y)\in X\times Y\subseteq\mathbb{R}^2$ and times $t\in [0,T]$ is characterized by the conditional intensity, defined as
\begin{align}
\label{eq:intensity}
    \lambda(x,y,t|\mathcal{H}_t) = \mu(x,y) + \sum_{\{k:t_k<t\}} g(t-t_k,x-x_k,y-y_k),    
\end{align}
where $\mathcal{H}_t = \{(x_i,y_i,t_i)\}_{i = 1}^n$ is the history of events up to time $t$ which we will omit for simplification of notation.
The background intensity $\mu(x,y)$ is often assumed to be time-independent while the triggering function $g(t-t_k,x-x_k,y-y_k)$ is generally chosen to be separable in time and space for computational simplicity.
For each event $(x_k,y_k,t_k)$, the number of offspring events follows a Poisson distribution with mean
\begin{align*}
    m = \int_{X\times Y}\int_T g(t,x,y) \text{d}t\text{d}(x,y).
\end{align*}
If properly normalized, $g(t-t_k,x-x_k,y-y_k)$ induces the probability distribution of the locations and times of these events.
After model fitting, the SEPP can be used to predict the locations and times of future events.
Assume we want to predict the number of events $N_{A,t}$ within an area $A\subseteq X\times Y$ at a given time $t = t'$.
This prediction can be obtained by computing the integral
\begin{align}
\label{eq:integrate_intensity}
    \widehat{N_{A,t'}} = \int_{A} \lambda(x,y,t|\mathcal{H}_{t'},t = t')\text{d}(x,y).
\end{align}

SEPP models were first applied to crime data for hot spot prediction by \citet{Mohler2011}. Initially, the authors suggested non-parametric estimation of $\mu$ and $g$ based on only background or offspring crimes respectively which requires a computationally expensive iterative stochastic declustering procedure. 
In subsequent work, \citet{Mohler2014} introduced a parametric approach that uses all data to estimate the background intensity with kernel density estimation and assumes a triggering function that is exponential in time and Gaussian in space.
The benefit of this parametric approach is that model parameters can be be estimated with a less expensive Expectation-Maximization procedure.
In field experiments with the Los Angeles Police Department and the Kent Police Department, United Kingdom, \cite{Mohler2015} forgo a complicated spatial model by fitting a cell-wise constant background intensity and a triggering function only exponential in time.

In this work, we draw on a fully parametric SEPP model that is inspired by the simulations conducted in \cite{Mohler2011}.
We assume a scaled Gaussian background intensity, defined as
\begin{align}
\label{eq:background_intensity}
    \mu(x,y) = \frac{\bar{\mu}}{2\pi(15)^2}\exp\left(-\frac{x^2}{2(15^2)}\right)\exp\left(-\frac{y^2}{2(15^2)}\right),
\end{align}
where the spatial deviation is chosen purposefully large to ensure support on the whole city map. 
Our triggering function is similar to the proposed parametric functions and takes the form
\begin{align}
\label{eq:triggering_function}
    g(t,x,y) = \theta\omega\exp(-\omega t )\exp\left(-\frac{x^2}{2\sigma_x^2}\right)\exp\left(-\frac{y^2}{2\sigma_y^2}\right).
\end{align}
Choosing a fully parametric model allows us to analyze a best-case scenario of the bias introduced by differential crime reporting rates as similar models can be used for data simulation and model fitting keeping error introduced by model misspecification at a minimum.
In addition, the model choice enables efficient computation of the prediction integrals in Equation~\ref{eq:integrate_intensity}.
For crime hot spot prediction, city maps are generally split into small areas by imposing a grid with fixed cell lengths. To predict the number of crimes within a cell at time $t$, integration over the estimated intensity function is necessary which can be computationally expensive depending on the exact model choice.
To the best of our knowledge, the model we use is similar to the model employed by PredPol's commercial hot spot prediction software.

\subsection{Expectation-Maximization Procedure}

The parameters of the SEPP model in Equation~\ref{eq:intensity}-\ref{eq:triggering_function} are estimated using maximum likelihood.
As an analytical solution is intractable, \cite{Veen2008} introduced an Expectation-Maximization (EM) algorithm that maximizes the log-likelihood.
Assuming we know the branching structure of the data set $\{(x_i,y_i,t_i\}_{i = 1}^n$, i.e. which events were triggered by which previous events and which events come from the background process, we introduce a latent variable $u_i$ which equals $j$ if crime $i$ was triggered by crime $j$ and $0$ if it was sampled from the background process.
Given these latent quantities, the complete-data log-likelihood of the parameter vector $\Theta = (\bar{\mu},\theta,\omega,\sigma_x,\sigma_y)$ can be written as
\begin{align*}
    l(\Theta) = &\sum_{i = 1}^n\mathbb{I}(u_i = 0)\log(\mu(x_i,y_i))\\
                &+ \sum_{i = 1}^n\sum_{j = 1}^n\mathbb{I}(u_i = j) \log(g(t_i-t_j, x_i - x_j, y_i - y_j))\\
                &-\int_{X\times Y}\int_T \lambda(x,y,t)\text{d}t\text{d}(x,y),
\end{align*}
where $\mathbb{I}(\cdot)$ is the indicator function.
Given a data set of crime events, the EM algorithm provides an iterative procedure of estimating the triggering probabilities $u_i$ and the parameters $\Theta$.
In the E step, we estimate the triggering probabilities based on current parameter values as
\begin{align*}
    \text{P}(u_i = j) = \begin{cases}\frac{g(t_i-t_j,x_i-x_j,y_i-y_j)}{\lambda(x_i,y_i,t_i)}&\text{ if }t_j<t_i,\\0&\text{ else.}\end{cases}
\end{align*}
and $\text{P}(u_i = 0) = \mu(x_i,y_i)/\lambda(x_i,y_i,t_i)$.
These latent values can then be plugged into the expected complete-data log-likelihood which gives
\begin{align*}
    \mathbb{E}[l(\Theta)] = &\sum_{i = 1}^n\text{P}(u_i = 0)\log(\mu(x_i,y_i))\\
                &+ \sum_{i = 1}^n\sum_{j = 1}^n\text{P}(u_i = j) \log(g(t_i-t_j, x_i - x_j, y_i - y_j))\\
                &-\int_{X\times Y}\int_T \lambda(x,y,t)\text{d}t\text{d}(x,y).
\end{align*}
In the M step, we maximize the expected log-likelihood with respect to $\Theta$ and return to the E step with the new parameter estimates.
This procedure is repeated until the parameter values converge.

\subsection{Bogot\'a Victimization and Reporting Survey}

Victimization rates---i.e. the fraction of the population who has been victim of a crime within a given time window---and victim crime reporting rates---i.e the fraction of crime victims who has reported the offense to the police---can generally not be assessed based on only police data but require large-scale surveys.
Often, these surveys are not conduced or published with a high-enough spatial resolution to give a sense of differences at a local level.
For instance, the US Bureau of Justice Statistics conducts a bi-annual National Crime Victimization Survey with around 95,000 households and publishes rates of victimization and crime reporting on a national level and aggregated by urban, suburban and rural areas \cite{morgan2018criminal}.

In order to study the effect of differential victim crime reporting on predictive policing systems, which are generally limited to a single city, a higher spatial resolution of victimization and crime reporting rates is required.
Fine-grained data sets like this are rare and, based on availability, we draw on district-level data from Bogot\'a, Colombia collected by Bogot\'a's chamber of commerce, C\'amara de Comercio de Bogot\'a (CCB). 

\begin{figure}
\centering
\begin{subfigure}{.49\columnwidth}
  \raggedright
  \vspace{2ex}
  \includegraphics[trim = 150 50 100 0, clip, width = 1\linewidth]{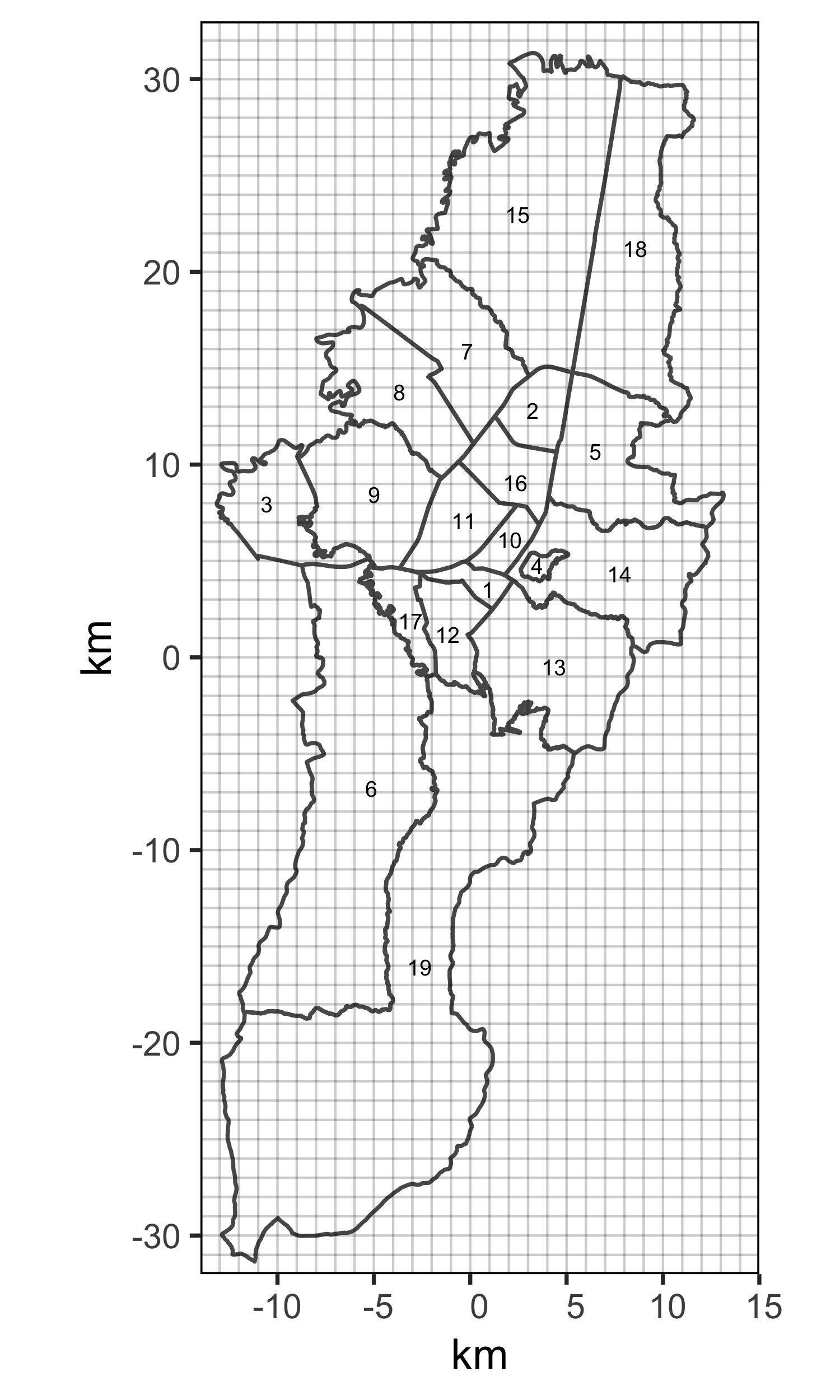}
\end{subfigure}%
\begin{subfigure}{.49\columnwidth}
  \raggedleft
  \scriptsize
  \setlength{\extrarowheight}{1px}
  \begin{tabular}{|C{0.1cm}|C{0.9cm}|C{0.7cm}|C{0.35cm}|C{0.35cm}|}
        \hline
        \textbf{Id} & \textbf{District} & \textbf{Pop.} & \textbf{Vict. rate} & \textbf{Rep. rate}\\ 
        \hline
        1 & Antonio Nari\~no & 109,176 & 15\,\% & 33\,\%\\ \hline
        2 & Barrios Unidos  & 243,465 & 12\,\% & 22\,\%\\ \hline
        3 & Bosa & 673,077 & 13\,\% & 26\,\%\\ \hline
        4 & Candelaria & 24,088 & 12\,\% & 22\,\%\\ \hline
        5 & Chapinero & 139,701 & 9\,\% & 28\,\%\\ \hline
        6 & Ciudad Bol\'ivar & 707,569 & 8\,\% & 17\,\%\\ \hline
        7 & Engativ\'a & 887,080 & 11\,\% & 20\,\%\\ \hline
        8 & Fontib\'on & 394,648 & 10\,\% & 19\,\% \\ \hline
        9 & Kennedy & 1,088,443 & 13\,\% & 28\,\%\\ \hline
        10 & Los M\'artires & 99,119 & 17\,\% & 25\,\% \\ \hline
        11 & Puente Aranda & 258,287 & 14\,\% & 32\,\% \\ \hline
        12 & Rafael Uribe Uribe & 374,246 & 12\,\% & 15\,\% \\ \hline
        13 & San Crist\'obal & 404,697 & 13\,\% & 21\,\%\\ \hline
        14 & Santa Fe & 110,048 & 17\,\% & 17\,\% \\ \hline
        15 & Suba & 1,218,513 & 5\,\% & 19\,\% \\ \hline
        16 & Teusaquillo & 153,025 & 14\,\% & 19\,\%  \\ \hline
        17 & Tunjuelito & 199,430 & 17\,\% & 23\,\% \\ \hline
        18 & Usaqu\'en & 501,999 & 18\,\% & 13\,\% \\ \hline
        19 & Usme & 457,302 & 9\,\% & 33\,\%\\ \hline
  \end{tabular}
\end{subfigure}
\caption{Bogot\'a district map with division into 1\,km$\times$1\,km grid cells for hot spot prediction and survey-based victimization and victim crime reporting rates. Districts differ notably in size, population numbers, and rates.}
\label{fig:bogota_rates}
\end{figure}

The bi-annual CCB crime perception and victimization survey includes approximately 10,000 randomly selected participants from all socio-economic statuses and all 19 urban districts of Bogot\'a.
Among other questions, participants are asked to indicate whether they have been the victims of a crime in the present calendar year and, if yes, whether they have reported the crime to the police. 
Results of the surveys are available on the CCB website and are used to inform the definition and adjustment of the city's public policies \cite{ccb_survey}. 
Not all of the published reports stratify results by districts.
For our experiments, we use victimization and victim crime reporting rates stratified by district based on the survey that covers the first half of 2014 \cite{ccb_pres}.
Districts, population sizes and rates are depicted in Figure~\ref{fig:bogota_rates}.
Both the crime victimization rates and the victim crime reporting rates vary significantly between different districts with victimization rates between 5\,\% and 18\,\% and victim crime reporting rate from 13\,\% to 33\,\%.
Although the range of both rates can be expected to vary significantly between different cities and countries,
this data allows us to analyze the impact of differential crime reporting on predictive policing in a realistic scenario.

\subsection{Synthetic Data Generation}

We simulate location and time of reported and unreported crime incidents in Bogot\'a districts according to the victimization and victim crime reporting rates displayed in Figure~\ref{fig:bogota_rates}.
In order to minimize possible errors due to model misspecification and instead concentrate on the effect of differential reporting rates, we sample data directly from a high-intensity SEPP
and subsample according to each district's victimization rate.
The background intensity of the SEPP is a sum over bivariate Gaussian distributions centered at 14 locations spread out evenly on the Bogot\'a map.
Each background crime triggers offspring according to a triggering function that is Gaussian in space and exponential in time coinciding with the model we are fitting (see Equation~\ref{eq:intensity}-\ref{eq:triggering_function}).
Since the data will be used to predict hot spots on a fixed grid, we impose a grid of 1\,km $\times$ 1\,km cells on the Bogot\'a map as depicted in Figure~\ref{fig:bogota_rates}.
District membership of each cell is decided based on its center and each point is attributed to the district of the cell it falls into. 
We dicretize the time component into daily units and simulate crime data for 2,190 timesteps (6 years) as follows:
\begin{enumerate}
    \item Sample a set of candidate points $\mathcal{C}=\{(x_i,y_i,t_i)\}_{i = 1}^{N}$
    from $\lambda$ and discard all points that fall outside of the city bounds or time horizon.
    \item For each district $d$ and data within its bounds $\mathcal{C}_d\subseteq\mathcal{C}$, we subsample
    $n_d\sim \text{Bin}(\lvert\mathcal{C}_d\rvert,p_d)$ of the points to form the true crime data set $\mathcal{D}$, where 
    $$
        p_d = \frac{\text{population}(d)\cdot \text{victimization rate}(d)\cdot 12 }{\lvert\mathcal{C}_d\rvert}.
    $$
    \item To get a data set of only reported crime, we subsample $n_d\sim \text{Bin}(\lvert\mathcal{D}_d\rvert, q_d)$ crimes for each district $d$ where $\mathcal{D}_d\subseteq\mathcal{D}$ is the set of crimes falling into the given district and
    $$
        q_d = \text{victim crime reporting rate}(d).
    $$
\end{enumerate}
We implicitly assume that each person is victim of at most one crime which leads to the time scaling factor $2190/(365/2) = 12$ in step 2 as the CCB survey provides rates of victimization for a half-year period.
In addition, district population counts are scaled by $1/40$ to speed up the run time of the whole simulation. 
The described sampling procedure for the true data $\mathcal{D}$ ensures that crime is sampled according to population size and victimization rates but remains distributed according to a thinned SEPP that can be accessed for evaluation of the ground truth conditional intensity. 
Since $\mathcal{D}\sim p_d\lambda$, the true expected number of crimes in a subarea $A_d$ of district $d$ in time $t=t'$ can be computed as
\begin{align*}
    \mathbb{E}[N_{A_d,t'}] = \int_{A_d} p_d \lambda(x,y,t|\mathcal{H}_{t'}, t = t')\text{d}(x,y),
\end{align*}
where $\mathcal{H}_{t'} = \{(x_i,y_i,t_i) \in \mathcal{C}: t_i<t'\}$.

Figure~\ref{fig:data} depicts a summary of the sampled number of crimes per district, the number of crimes expected according to above integral and the number of crimes as implied by the CCB survey showing that the synthetic data set has the desired rates of victimization for each of the districts.

\section{Results}

\subsection{Hot spot prediction procedure}
We fit SEPP models (see Equation~\ref{eq:intensity}-\ref{eq:triggering_function}) on the full and reported crime data by discarding the data from the first 500 simulated time steps and training on the subsequent 1,500 days ($\approx$ 4 years) of sampled incidents. 
Ignoring the first 500 time steps omits the period in which the data generating SEPP is converging to its equilibrium rate and provides a data set that more closely resembles the crime data over fixed time windows we would expect to see in practice.
In addition, the time range of approximately 4 years is reasonably close to real crime data sets and falls well within the range of 2-5 years that is suggested by PredPol specifically \cite{predpol_website_years}.

The fitted models are used to predict crime intensities on a day-to-day basis for 189 evaluation days where, after each time step, the data for the time step is observed and added to the estimated intensity function for future predictions.
On each prediction day, we compute the models' intensity integrals in each of the 1\,km $\times$ 1\,km Bogot\'a grid cells.
These integrals correspond to the absolute predicted crimes per cell and are subsequently used for hot spot selection.
Since police are generally only able to patrol small fractions of a city effectively, we select the top 50 cells with highest predicted crime as hot spots which corresponds to approximately 5.7\,\% of the city's area.
Results are aggregated over 50 simulation runs where each simulation samples a new crime data set.

\subsection{Equity between districts}
\subsubsection{Relative number of predicted hot spots}

\begin{figure*}
    \centering
    \includegraphics[scale = 0.14]{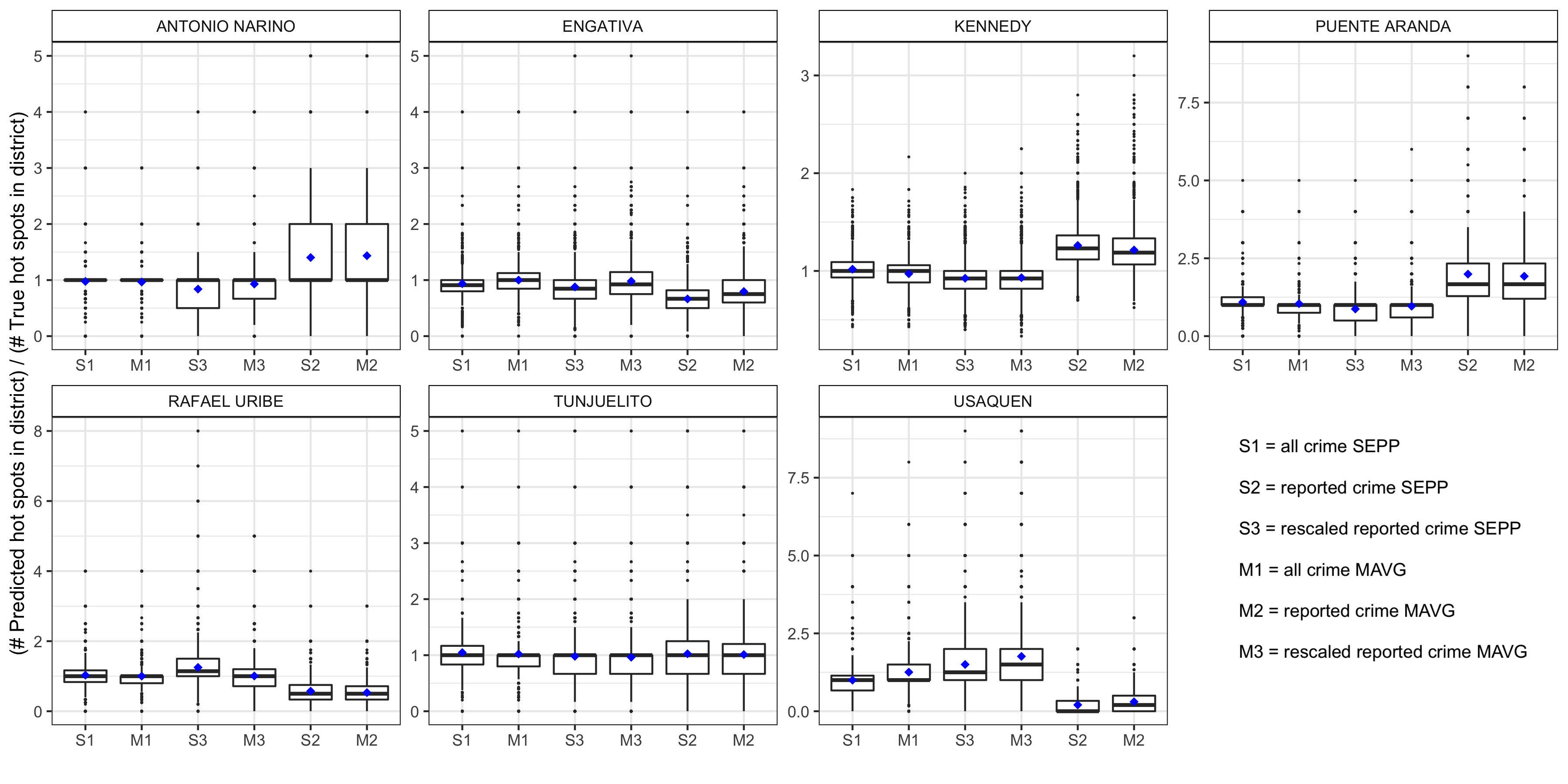}
    \caption{Relative number of predicted crime hot spots for a selection of Bogot\'a districts. Each data point represents a district-specific fraction at a given evaluation time step (189 days) in a given simulation run (50 runs). A total of 50 hot spots are selected at each time step. See Figure~\ref{fig:rel_count_appendix} for relative predicted hot spot counts for all districts.}
    \label{fig:rel_count_box}
\end{figure*}

We now discuss the equity of hot spot selection at a district-level. We start by examining how the number of predicted hot spots compares to the number of true hot spots per prediction day in each district.
In the case where police are deployed in accordance with the model's predictions, this
directly corresponds to the degree of police presence per district relative to a best-case hot spot policing program in which the true crime distribution is known.

\begin{figure}
    \centering
    \includegraphics[scale = 0.13]{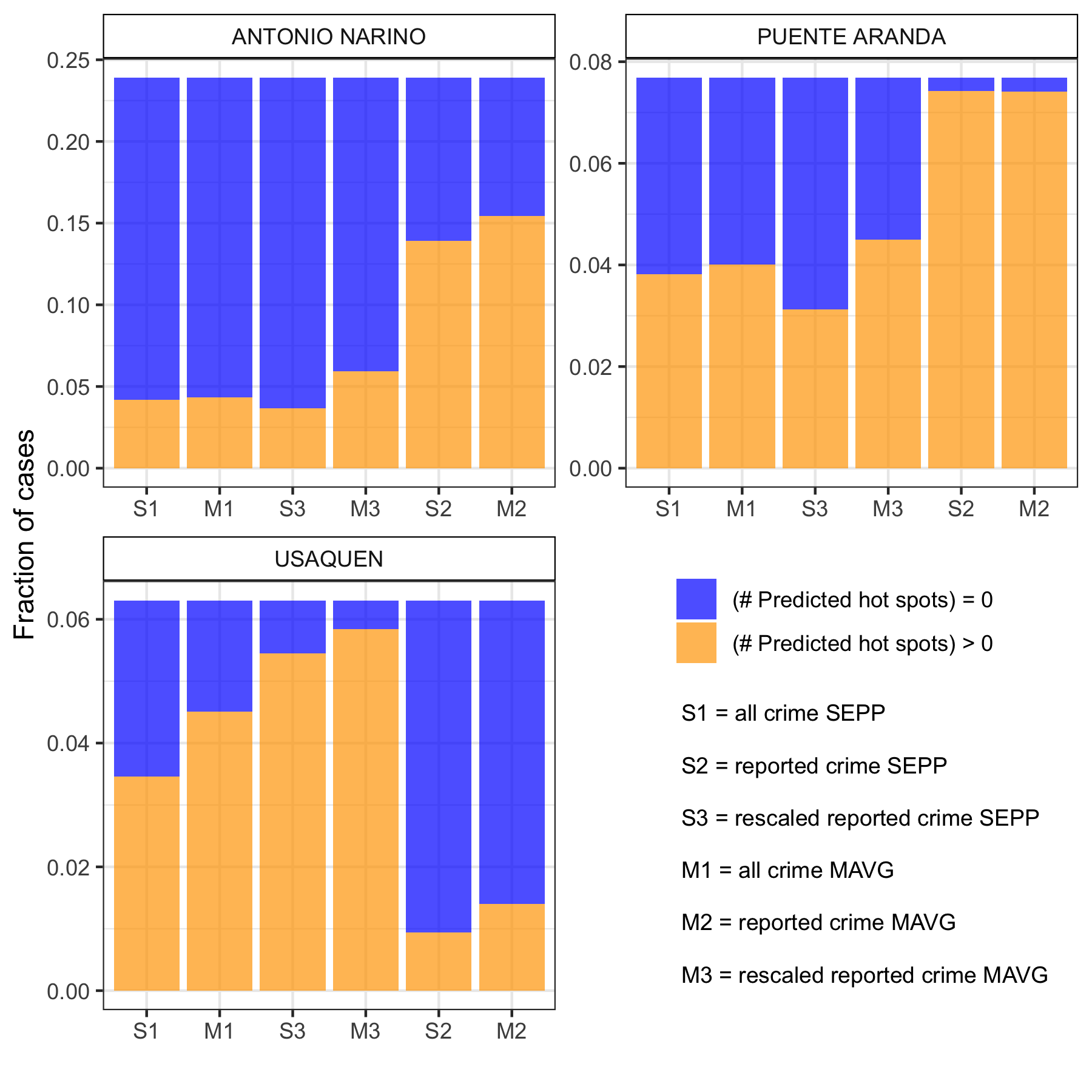}
    \caption{Fraction of prediction time steps with no true hot spots in districts. We separate instances into cases with predicted and no predicted hot spots. 
    See Figure~\ref{fig:rel_count_no_true_appendix} for a version with all districts.}
    \label{fig:rel_count_small}
\end{figure}

Figures~\ref{fig:rel_count_box} and \ref{fig:rel_count_small} depict the relative hot spot counts for a subset of districts over all evaluation time steps and simulation runs.
For Figure~\ref{fig:rel_count_box}, we set the relative count to $1$ for cases in which the district has zero true hot spots and the model correctly predicts zero hot spots and exclude cases with zero true but non-zero predicted hot spots.
We see that the SEPP model that was trained on all crime data, i.e. reported and unreported, performs well at selecting the correct number of hot spots uniformly over all districts (S1).
This observation is unsurprising given that the fitted model closely resembles the data generating model.

In contrast, the SEPP model that was trained on only reported crime data (S2) is found to have differential performance across districts.
Although in some districts, e.g. in Tunjuelito, the relative hot spot counts of the two models appear to be similar, the model with under-reporting on average overestimates the number of hot spots in districts such as Antonio Nari\~no, Puente Aranda or Kennedy, while underestimating the number of hot spots in districts such as Usaqu\'en, Rafael Uribe Uribe or Engativ\'a. 
The direction of the introduced error aligns with the victim crime reporting rates of the respective districts as compared to a Bogot\'a-wide average with fewer of the true hot spots detected in low reporting areas and instead overly many hot spots predicted in high reporting areas.
In Usaqu\'en, which with 13\,\% has the lowest victim crime reporting rate among all districts, only 20.4\,\% of the number of true hot spots are predicted on average.
Meanwhile in Kennedy, which has a comparatively high reporting rate of 28\,\%, the model on average predicts 126.1\,\% the number of true hot spots.

Thus far, we have disregarded cases in which none of the true hot spots fall into a given district but the prediction model selects one or more cells.
Figure~\ref{fig:rel_count_small} gives a summary of the fraction of cases with no true hot spots, further confirming the observed displacement effect of hot spot predictions.
In Usaqu\'en, the number of times crime hot spots are predicted when none of the true top 50 crime hot spots lie in the district is over twice as high in the full data SEPP compared to the reported crime SEPP. 
The same fraction increases more than threefold in the high-reporting district Antonio Nari\~no, and almost twofold in Puente Aranda.
Notably, Figure~\ref{fig:rel_count_small} also shows that the displacement effect both impacts districts that almost always have areas with highly concentrated crime and districts that do not.  This phenomenon is a function of victimization rates, population sizes and the size of districts. 

Finally, average absolute numbers of over- or underpredicted hot spots are displayed in Figure~\ref{fig:diff_count_appendix}. 
Although comparison of relative counts ensures that districts of different sizes are evaluated similarly, in some cases we might be interested in the number of grid cells affected by the introduced bias as they roughly relate to the number of impacted individuals. 
For example, we see that the displacement of predicted hot spots based on differential victim crime reporting rates leads to on average 3.3 too many hot spots predicted in Kennedy while only 0.64 too many cells in Antonio Nari\~no are selected on average. 

Overall, differential reporting rates across districts seem to lead to differentially well-measured aggregate crime levels which distorts the distribution of hot spots. 
If the police follows the model's recommendations, the consequence would be an unfair allocation of police patrols where areas with low victim crime reporting rates are met with artificially decreased police presence while areas with higher reporting rates are chronically over-policed.

\subsubsection{Crime threshold for hot spot selection}

Calculating relative counts of predicted hot spots gives insights into how much under- or over-policing we can expect per district.
A natural way of comparing between districts is to look at the true crime rates required for a cell to be selected as a hot spot.
If this threshold is much lower for some districts than for others, the consequence could be more average police presence in these districts despite similar or even higher crime levels in other areas. 

\begin{figure}
    \raggedright
    \begin{subfigure}{0.69\columnwidth}
    \vspace{2ex}
    \includegraphics[trim = 250 0 150 0, clip , scale = 0.15]{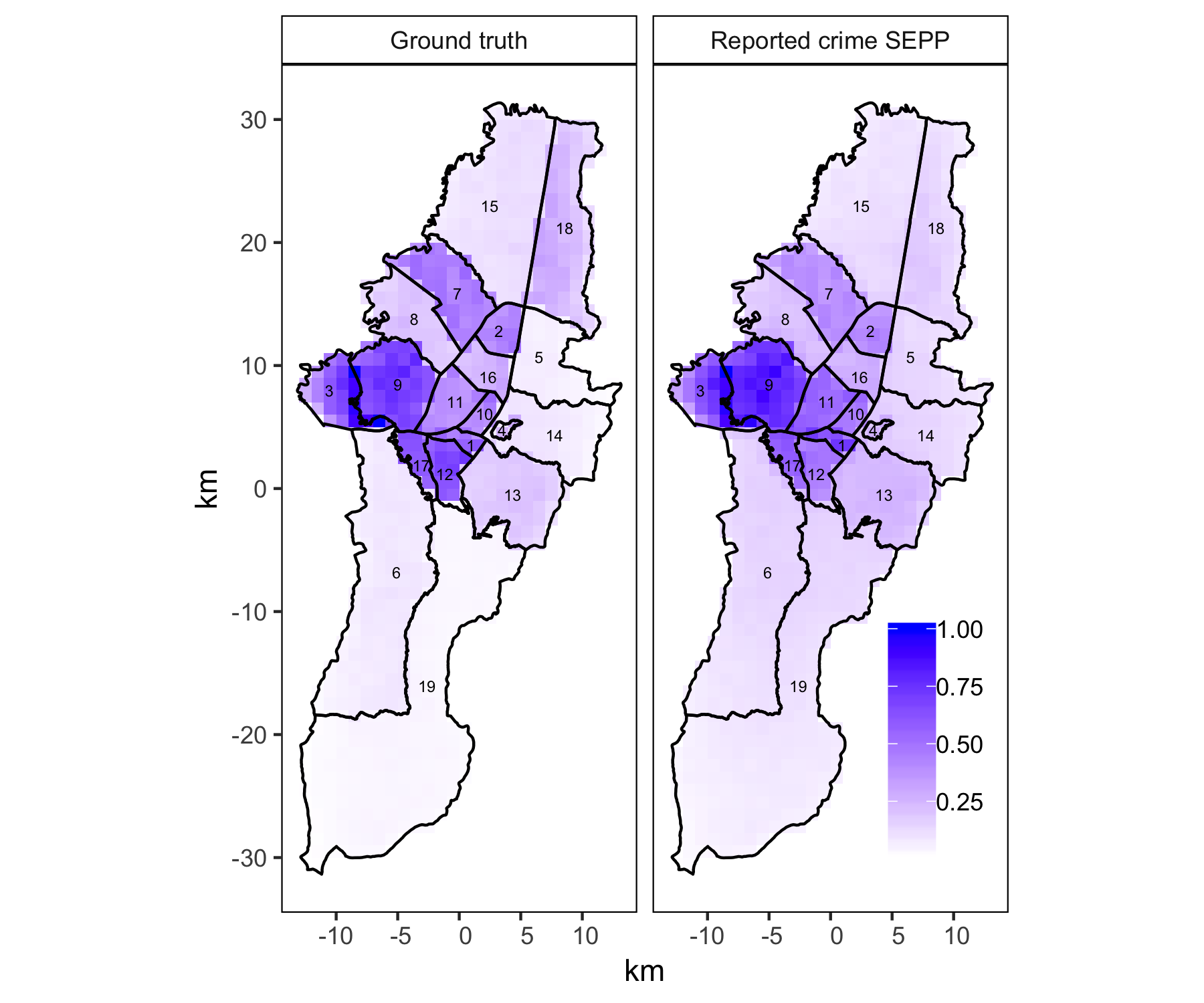}
    \end{subfigure}
    \begin{subfigure}{0.3\columnwidth}
        \raggedleft
  \scriptsize
  \setlength{\extrarowheight}{1px}
  \begin{tabular}{|C{0.1cm}|C{0.9cm}|}
        \hline
        \textbf{Id} & \textbf{District}\\ 
        \hline
        1 & Antonio Nari\~no\\ \hline
        2 & Barrios Unidos\\ \hline
        3 & Bosa\\ \hline
        4 & Candelaria\\ \hline
        5 & Chapinero\\ \hline
        6 & Ciudad Bol\'ivar\\ \hline
        7 & Engativ\'a\\ \hline
        8 & Fontib\'on\\ \hline
        9 & Kennedy\\ \hline
        10 & Los M\'artires\\ \hline
        11 & Puente Aranda\\ \hline
        12 & Rafael Uribe Uribe\\ \hline
        13 & San Crist\'obal\\ \hline
        14 & Santa Fe\\ \hline
        15 & Suba\\ \hline
        16 & Teusaquillo\\ \hline
        17 & Tunjuelito\\ \hline
        18 & Usaqu\'en\\ \hline
        19 & Usme\\ \hline
  \end{tabular}
    \end{subfigure}
    \caption{Normalized average crime in each cell. The left side depicts the average over true intensity integrals, while the right side uses predictions from the SEPP model trained on only reported crime data. In both cases, we normalize by dividing by the respective maximum average prediction value.}
    \label{fig:heatmap}
\end{figure}

Figure~\ref{fig:heatmap} shows that the predicted crime rates implied by the reporting data SEPP model present a differentially well-adjusted approximation of true crime rates.
Comparing the normalized maps in the Figure, the reported crime SEPP appears to overestimate the relative concentration of crimes in the high-reporting regions Kennedy and Antonio Nari\~no, and underestimate the relative concentration of crimes in low-reporting districts such as Rafael Uribe Uribe and Usaqu\'en. 
Moreover, crime rate prediction seems to perform poorly in areas with little true crime.
While the ground truth shows clear differences between crime intensities in areas such as Ciudad Bol\'ivar and Usme, the model predictions in these districts appear to be almost indistinguishable.

\begin{figure*}
    \centering
    \includegraphics[scale = 0.14]{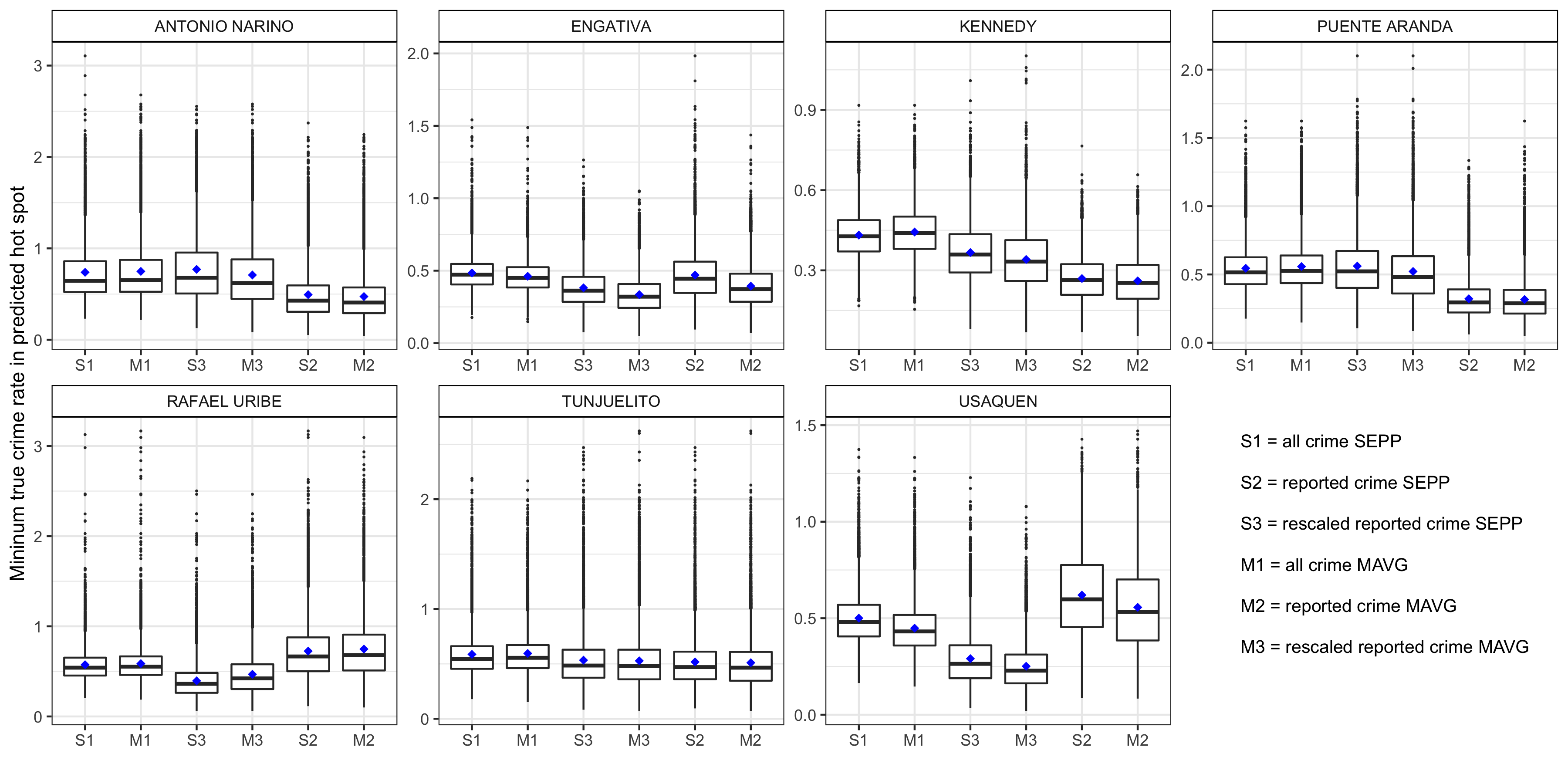}
    \caption{True crime thresholds for hot spot selection in a set of Bogot\'a districts.
    Each point corresponds to an evaluation time step (189 days) and a simulation run (50 runs). See Figure~\ref{fig:rel_min_cg_appendix} for all districts.}
    \label{fig:min_cg}
\end{figure*}

In order to measure equity of model predictions between districts, we consider the minimum true crime rate that leads to a predicted hot spot at each prediction step and summarize the results in Figure~\ref{fig:min_cg}.
Since exact crime counts vary over time and this metric omits steps with no predicted hot spots falling into the respective district, the average thresholds have some variability even for full data models.
However for districts that are regularly predicted to have hot spots, the full data SEPP model (S1) exhibits very similar hot spot prediction threshold of around 0.5 expected crimes per cell and time step where the low threshold is explained by the population scaling we conducted while simulating Bogot\'a crime data. 
In contrast, the model trained on only reported crime data results in varying thresholds even across districts which are regularly predicted to have hot spots. 
The district-wide average threshold of 0.45 true expected crimes per cell is increased in areas with low crime reporting, e.g. to a rate of 0.73 true crimes on average in Rafael Uribe Uribe and 0.62 in Usaqu\'en.
At the higher end of victim crime reporting rates, grid cells in Puente Aranda on average only require a rate of 0.32 true crimes and cells in Kennedy only 0.27 to be selected as a crime hot spots.
More concretely, this means that on average the minimum true crime rate that leads to a predicted hot spot in Rafael Uribe Uribe is 2.7 times the minimum crime rate required in Kennedy.
In order to rule out the possibility that Kennedy's threshold is artificially high because all of the cells in the district are regularly selected as hot spot, we examine the absolute predicted hot spot counts and find that at no time step more than 72.97\,\% of Kennedy is selected as hot spot area with a mean of 48.18\,\%.

These findings imply that crime hot spot prediction in real-world settings with differently sized regions and differential victimization and crime reporting rates can have noticeably biased outcomes that lead to over-policing of some areas of a city while others have higher levels of crime.

\subsection{Scaling by victim crime reporting rates}

It is not unusual for police and predictive policing algorithms to leverage data sets beyond registered crime incidents \cite{Shapiro2017,Jefferson2017,GimnezSantana2018}.
In the case presented here, one could imagine pairing the reported crime data with the survey data to attempt to correct the bias introduced by differential crime under-reporting.
Intuitively, this entails rescaling the predicted crime rates according to the reporting rates. Of course, in most cases exact crime reporting rates are unknown to the police. However, as we discuss in this section, even in cases where the reporting rates are known, this rescaling does not necessarily recover the true crime distribution at the finest level. 

We explore the rescaling approach as an additional model in our hot spot prediction simulation by taking the integrated intensities in grid cells supplied by the reporting data SEPP and dividing them by the victim crime reporting rate of the respective district.
After rescaling, we select the cells with the top 50 highest predictions as hot spots analogous to the other models.
The relative predicted hot spot counts of the rescaled model (S3) are displayed along the other models in Figure~\ref{fig:rel_count_box}.
Across the displayed districts, the mean relative number of predicted hot spots is just as close or closer to the number predicted by the full data model (S1) than the reporting data based predictions (S2).  This indicates that the rescaling strategy successfully reduces outcome disparities.
However, this conclusion is called into question when examining the implied minimum true crime rate for hot spot selection shown in Figure~\ref{fig:min_cg}.
For example in Usaqu\'en, the rescaled model implies a visibly lower average true crime threshold for hot spot selection than the full data model, and in Engativ\'a the difference between the full data and rescaled models appears to be larger than the difference between the full and reporting data models.

The conflict between the equity measures is observed because the relative predicted hot spot counts are an aggregate metric over all cells and not sensitive to which cells are selected, in contrast to the minimum true crime threshold for hot spot selection.
Rescaling of the reporting data SEPP predictions increases predictions in all cells of a district by the same factor without accounting for how much crime was unobserved in each of the cells. 
As a consequence, the rescaled model selects an approximately correct number of hot spots in many of the districts while the exact cells might not coincide with the true hot spots.
In order to recover the cell-wise true crime distribution, a cell-by-cell rate of crime reporting would be required,
which presupposes separate representative surveys in hundreds of micro-areas. While incorporating victimization survey information does help to reduce disparities to some extent, it evidently does not suffice in order to fully debias the prediction system.

\subsection{Comparison to a moving average model}

In this section we study the behavior of a simple moving average (MAVG) prediction model to assess whether our findings hold more generally outside of the SEPP prediction model setting. MAVG prediction models are fitted analogously to the SEPP models on the full and reported crime data sets. 
Despite being perhaps the simplest possible prediction model, MAVG's have been found to perform particularly well for detecting long-term hot spots \cite{Fitzpatrick2018} in real world data.

For our application, we aggregate crime in the same 1\,km $\times$ 1\,km grid cells previously used and fit a within-cell MAVG model to predict the daily crime counts on the same training data sets as before. 
To obviate the need for tie-breaking that would arise if using simple averaging, we instead employ an exponentially-weighted MAVG model. 
The same model parameter is estimated for the entire Bogot\'a grid by searching over a linear scale of bandwidths for the exponential smoothing kernel and selecting the parameter that induces minimal average error with lagged prediction on the training data set. 
The models are updated on a daily basis by incorporating the crime counts of the previous day.

The performance of the full data MAVG model (M1), the reporting data MAVG model (M2), and the rescaled MAVG model (M3) are depicted in Figure~\ref{fig:rel_count_box}, \ref{fig:rel_count_small} and \ref{fig:min_cg} alongside the corresponding SEPP models. 
We observe that the MAVG models generally perform similarly to their respective SEPP counterparts.  The full MAVG model (M1), for instance,  performs on par with the full data SEPP model.
Likewise, the reporting data MAVG (M2) induces similar outcome disparities in relative hot spot counts and minimum true intensities as the SEPP trained on victim crime reporting data, and the rescaled MAVG model (M3) struggles to correct the finer resolution bias similarly to the rescaled SEPP model.

At first glance these similarities might be surprising, 
especially because the true data was simulated from a SEPP.
However both the SEPP and the MAVG model follow similar modeling ideas. 
While the MAVG model forgoes the spatial modeling component by discretizing into grid cells prior to prediction, whereas the SEPP has a continuous underlying intensity that is later integrated over grid cells, both methods model the time between events with an exponential function.
In addition, both models make predictions based on a weighted average of previous nearby events and the weights can be fairly similar if we assume that the spatial deviation of the triggering function is small in comparison to the size of the grid cell such that most offspring crimes fall into the same cell as their parent.
This assumption is often justified as the criminology literature tends to describe crime hot spots as micro areas of only a few blocks or street segments with high concentration of crime \cite{Fitzpatrick2019}. 
Indeed, in their randomized controlled field trials, the researchers affiliated to PredPol omit the spatial component of the SEPP altogether and discretize crimes into cells before modeling \cite{Mohler2015}.

\section{Discussion}

Our analysis demonstrates how predictive policing systems exclusively trained on victim crime reporting data can lead to spatially biased outcomes due to geographic heterogeneity in crime reporting rates.
This in turn can result in over-policing of certain communities while others remain under-served by police.

Our findings are based on synthetic crime data simulated according to district-level victimization and victim crime reporting rates published by Bogot\'a's chamber of commerce, C\'amara de Comercio de Bogot\'a.
We empirically evaluate the equity of predictions across districts of a hot spot prediction algorithm similar to the models used by PredPol.
Our findings suggest that districts with low crime reporting rates have fewer of their crime hot spots detected by the algorithm.
Conversely, districts with high crime reporting rates are found to have a higher concentration of predicted hot spots than the true crime levels would justify. 
Moreover, the effective true level of crime required for the model to predict a hot spot is found to vary by more than a factor of two across the districts.

We explore if known victim crime reporting rates can be used to debias hot spot predictions by scaling crime expectations appropriately. 
The results suggest that this is unsuccessful when reporting rates are known at a district level but hot spots are predicted at a smaller individual cell level since noise introduced by individually thinned crimes is propagated to the rescaled predictions which makes singling out of specific cells in comparison to other cells in the same district difficult.

Prior work has focused on feedback loops and the potential harms of arrest data-based predictive policing systems \cite{Ensign2018,Lum2016}.  Yet, in practice, predictive policing systems are based on data from victim crime reports \cite{cheetham_2019}.  Our work presents an initial step toward understanding the effect of bias in victim crime reporting data on predictive policing systems.  Our analysis demonstrates the importance of considering reporting rate variation when assessing predictive policing systems for potential harms and disparate impacts.  

\subsection{Limitations}
\subsubsection{Crime location vs. survey location}
Victimization surveys generally provide us with information on crime reporting based on where people live, not based on where crimes occur. 
On a small scale like a single city, this spatial disparity makes it hard to take survey-based information into account for police allocation.  While this limitation of how survey data is collected does not invalidate our findings---in our simulations we treat the reporting rates as reflecting rates of reporting for crimes \textit{occurring} in the given district---it does present a challenge for using such survey data to de-bias predictions in practice.  In order for survey data to be useful for this purpose, questions need to ask not only where respondents reside, but also where the victimization(s) occurred.  

\subsubsection{Static reporting rates and potential deterrence effects}
Thus far, we do not take the effects of the actual interventions in the form of patrolled hot spots into account. 
We hypothesize that both victimization rates and victim crime reporting rates can be susceptible to police presence and a model that jointly describes the interplay of crime, reporting rates and police deployment is required for a more complete picture. 
One component currently omitted is a deterrence effect of policing.
Failing to consider such effects could result in the reallocation of police patrols away from neighbourhoods where they are having the intended deterrence effect, precisely because reported crime rates would be lower when police are successful in deterring crime.

\subsection{Implications and Generalizability}

\subsubsection{Relationship to socioeconomic advantage.}

Research on victim crime reporting shows that the decision to report a crime is influenced by the severity of crime \cite[e.g.][]{Greenberg1992,Goudriaan2006}, victim characteristics \cite[e.g.][]{Slocum2017,Hullenaar2020}, and contextual factors such as neighborhood characteristics \cite[e.g.][]{baumer2002neighborhood,slocum2010,zhang2007}.
While we lack information on victim and crime characteristics in the survey data, we are able to speak to a number of socio-technical implications of our results.

Prior research finds links between severe socioeconomic neighborhood disadvantage and lower reporting rates for simple assault incidents \cite{baumer2002neighborhood}.
\citet{Goudriaan2006} obtain similar results analysing crime incidents from the Netherlands paired with the Dutch Police Population Monitor survey.
Some studies describe a more indirect effect of socioeconomic status on the likelihood of reporting.
For example, \cite{Berg2011} find that victims who are involved in illegal behavior are less likely to report violent acts against them to the police, and this effect is particularly pronounced in disadvantaged neighborhoods.
The findings of \cite{Slocum2017} suggest that prior police-initiated contact with law enforcement has a negative impact on the reporting of future crime that is amplified for African Americans and poorer individuals.
The authors of \cite{Cattaneo2008} study the help-seeking behavior of women who experience intimate partner violence. The study finds that, for the lowest income women, the severity of violence does not predict whether law enforcement is contacted.
With increasing income the severity becomes more indicative of victim crime reporting.

In the Bogot\'a survey data, there appears to be no clear association between reporting rates and socioeconomic advantage at the district level.  
Ciudad Bol\'ivar, a district with large urban slums that is home to some of the most socioeconomically disadvantaged residents of Bogot\'a, has a reporting rate of 17\%.
In line with previous research, this lies well below the average reporting rate across districts of 22.7\%.
However, Usaqu\'en, the district with the lowest reporting rate of 13\%, is also one of the wealthiest districts in Bogot\'a.
We hypothesize that this is in part explained by the spatial clustering of specific crime types. In particular, Usaqu\'en experiences a greater proportion of residential burglary and theft than other districts \cite{ccb_report,GimnezSantana2018}.
Given that victim crime reporting rates vary based on perceived severity \cite{Xie2019review}, this might contribute to a decreased victim crime reporting rate. Additionally, this may also be influenced by intra-district heterogeneity of wealth, as socioeconomically disadvantaged neighborhoods such as El Codito are also located in this district. 

There is no simple relationship between socioeconomic level of districts and the geographical disparities induced by the hot spot prediction algorithm.
This is in part driven by the observed complexity in the relationship between socioeconomic status and crime reporting at the district level. 
For example, areas that we project to be over-policed under hot spot policing include the middle class district Antonio Nari\~no, the lower middle or working class district Puente Aranda and the working and lower class district Kennedy.
Areas observed to be under-predicted likewise include districts inhabited by upper, middle, working and lower class residents.  Thus our findings \textit{do not} indicate that variation in crime reporting rates systematically disadvantages Bogot\'a's districts in a manner that falls along socioeconomic lines.   

\subsubsection{Relationship to demographics.} Demographic factors such as age \cite{baumer2002neighborhood,bosick2012reporting,hashima1999violent,watkins2005examining}, gender \cite{baumer2010reporting} and race \cite{xie2012racial,avakame1999did} can play a role in victim crime reporting.
\citet{Desmond2016} examine the change in Milwaukee's crime reporting rates after public broadcast of police violence against an unarmed black man. They find that, particularly in black neighbourhoods,  residents were far less likely to report crime to police following the incident.
Ultimately, race and ethnicity are often correlated with socioeconomic status and location, which makes it difficult to identify the direct relationships between demographic variables and victim crime reporting rates \cite{Shapiro2017}. 

Due to data limitations, we are unable to provide an indepth discussion of the relationship between race, ethnicity and predictive disparities in the Bogot\'a context, as we do not have access to demographic information on the victimization survey participants. A discussion of the Bogot\'a-specific interplay of race, ethnicity, crime and policing, and how it might generalize to other contexts, thus remains beyond the scope of this paper.

\subsubsection{Generalizability to other jurisdictions.}
Crime reporting decisions also operate in a macrolevel context encompassing specific cities, local governments or whole nations \cite{Xie2019review}.
\citet{Gutierrez2015} find that, within the US, metropolitan areas with greater proportions of foreign-born or non-US citizens have decreased crime reporting rates, and 
the results of \cite{Miller2018} suggest that cities with more female police officers have higher rates of victim crime reporting for violent crimes against women.
\citet{Goudriaan2004} analyze data from 16 Western industrialized countries and find that differences in crime reporting rates are not entirely explained by crime types, individual and local contexts, but vary with nation-level factors such as the perceived competence of the police at large.

Since our study is exclusively based on survey data from Bogot\'a, specific findings do not necessarily generalize to other geographies.  
In particular, while our analysis did not find evidence of a simple relationship between socioeconomic factors at a district level and predictive disparities, results would likely be different in regions---or at resolutions---where socioeconomic factors are more directly associated with reporting rates.

Although the specific spatial distribution and societal implications of the predictive disparities are likely to vary between different jurisdictions, our results suggest that some form of outcome disparity can be expected if victim crime reporting rates have sufficient spatial variation.
Such spatial variation is relatively commonplace and can be expected if, for example, the city has some amount of socioeconomic segregation since crime reporting rates vary with neighborhood disadvantage \cite{Xie2019review,baumer2002neighborhood,Berg2011}.

\subsubsection{Combining data sources and debiasing}

Predictive policing algorithms rely on crime data collected by law enforcement that has repeatedly been found to be flawed, biased or in other ways `dirty’ \cite{Richardson2019}. Much of the attention has focused on biases in police-initiated and particularly arrest data, for instance racial bias in drug related arrests or traffic stops in the US \cite{Pierson2020,BECKETT2006}.
PredPol acknowledges some of these biases and  publicly states that no drug-related offenses, traffic stops or arrest data are used in their prediction system \cite{predpol_2017}.  
Yet there is a lack of transparency as to how the data types that are `too biased’ to be included were identified, to what extend other data sources are biased, and which types of biases were considered. 
To the best of our knowledge, there has been no consideration of
reporting biases although their link to
socioeconomic, demographic and cultural factors as described in earlier sections is known. 
Motivated by this problem, we show that, even when predictive policing algorithms only operate on victim crime reporting data, and thereby attenuate the effects of biased police arrest practices, differential victim crime reporting rates can lead to geographically biased prediction outcomes. 

In addition to police data, some predictive policing systems incorporate contextual data from other sources \cite{Shapiro2017,GimnezSantana2018}. For example, HunchLab combines public reports  of crime and requests of police assistance with data including weather patterns, geographical features, schedules of major events and even moon phases \cite{Shapiro2017}.
In the setting of the hot spot prediction analyzed in this study, one could imagine the proposal to account for the bias introduced by differential reporting rates by scaling model outputs by the survey-based geographically stratified reporting rates.
However, our preliminary experimentation suggests that, although in some cases bias can be decreased, a complete mitigation is not possible if the surveyed victim crime reporting rates do not have sufficient spatial resolution. For successful debiasing, we would require close-to-optimal estimates of victim crime reporting rates at the grid cell level, which is impossible to obtain in practice. Ultimately, it is unclear if debiasing victim crime reporting data is any easier than the unsuccessful previous efforts of mitigating bias introduced by arrest data.

\clearpage

\bibliographystyle{ACM-Reference-Format}
\bibliography{literature}

\clearpage
\appendix
\section{Supplementary figures}

\begin{figure*}[b]
 \vspace{-2ex}
    \centering
    \includegraphics[scale = 0.13]{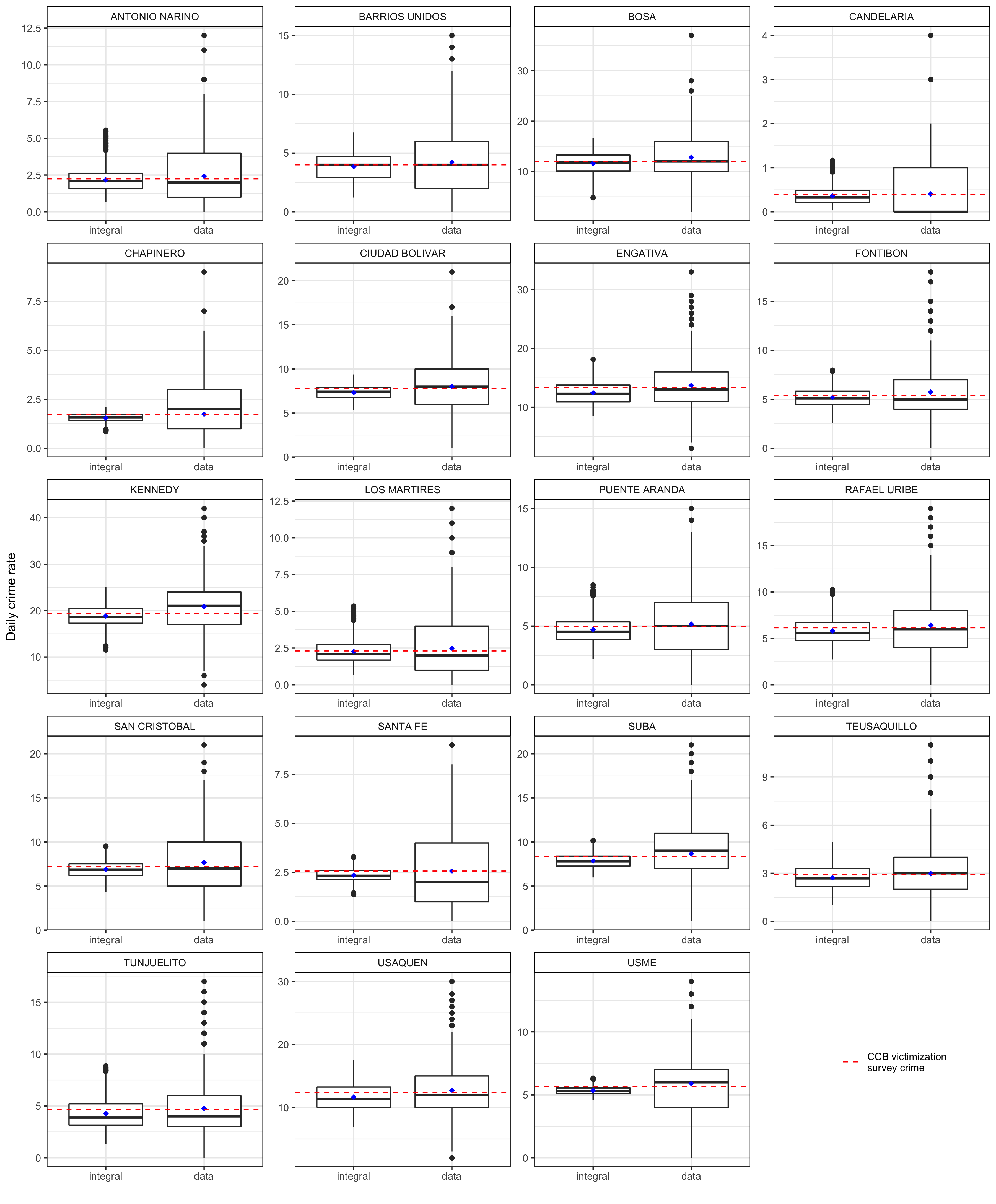}
    \caption{District-wise sanity check of synthetic crime data over all 2,190 time steps. The average daily counts of simulated data align well with the rates obtained by integration of the data generating thinned SEPP and the desired rates implied by the CCB victimization survey.}
    \label{fig:data}
\end{figure*}

\begin{figure*}
    \centering
    \includegraphics[scale = 0.13]{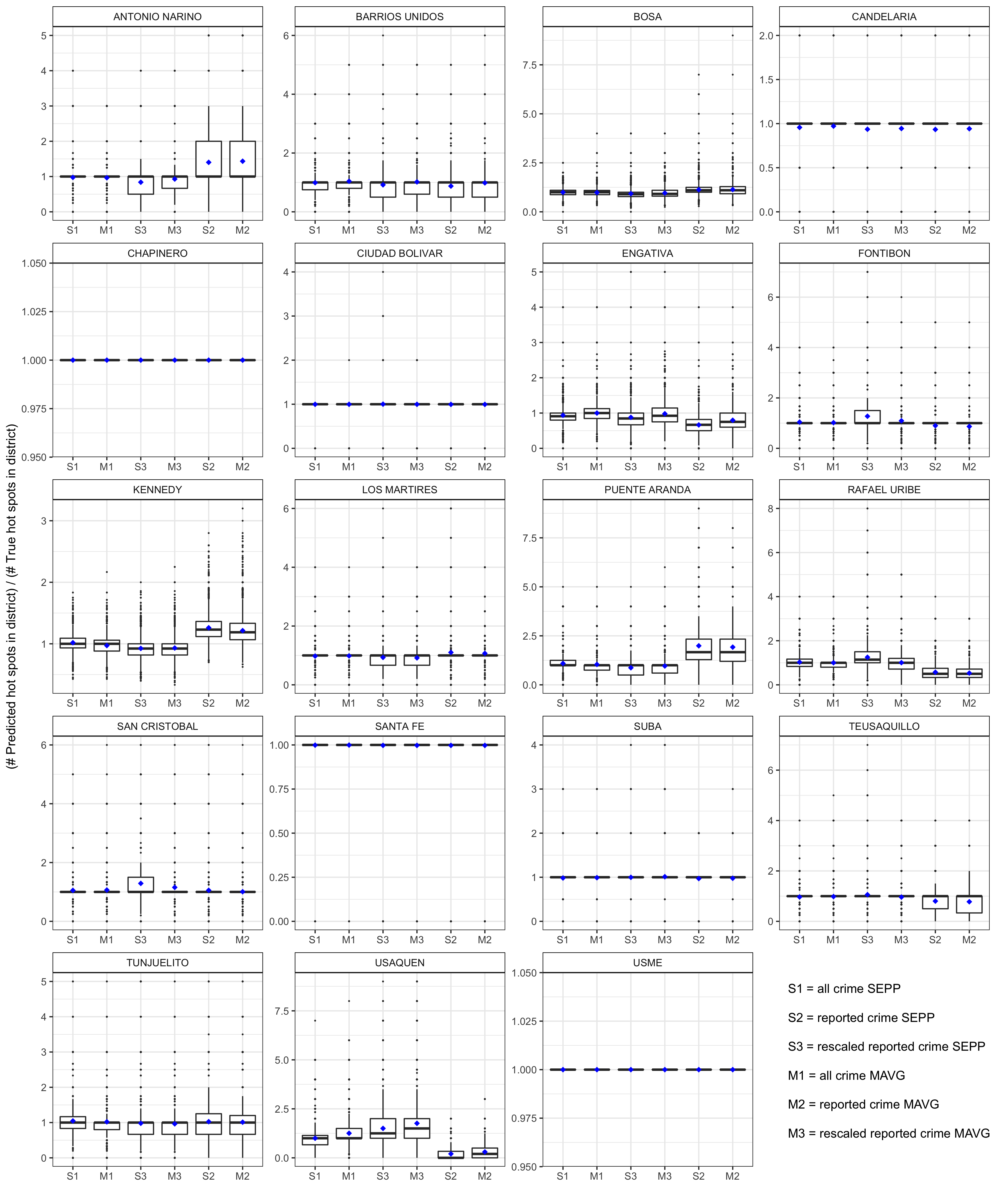}
    \caption{Relative number of predicted crime hot spots in Bogot\'a districts. Each data point represents a district-specific fraction at a given evaluation time step (189 days) in a given simulation run (50 runs). A total of 50 hot spots are selected at each time step. If both the true and predicted number of hot spots is zero, we set the relative count to one. Cases for which the number of predicted hot spots is non-zero but no true hot spots are available are excluded for visualization.}
    \label{fig:rel_count_appendix}
\end{figure*}

\begin{figure*}
    \centering
    \includegraphics[scale = 0.13]{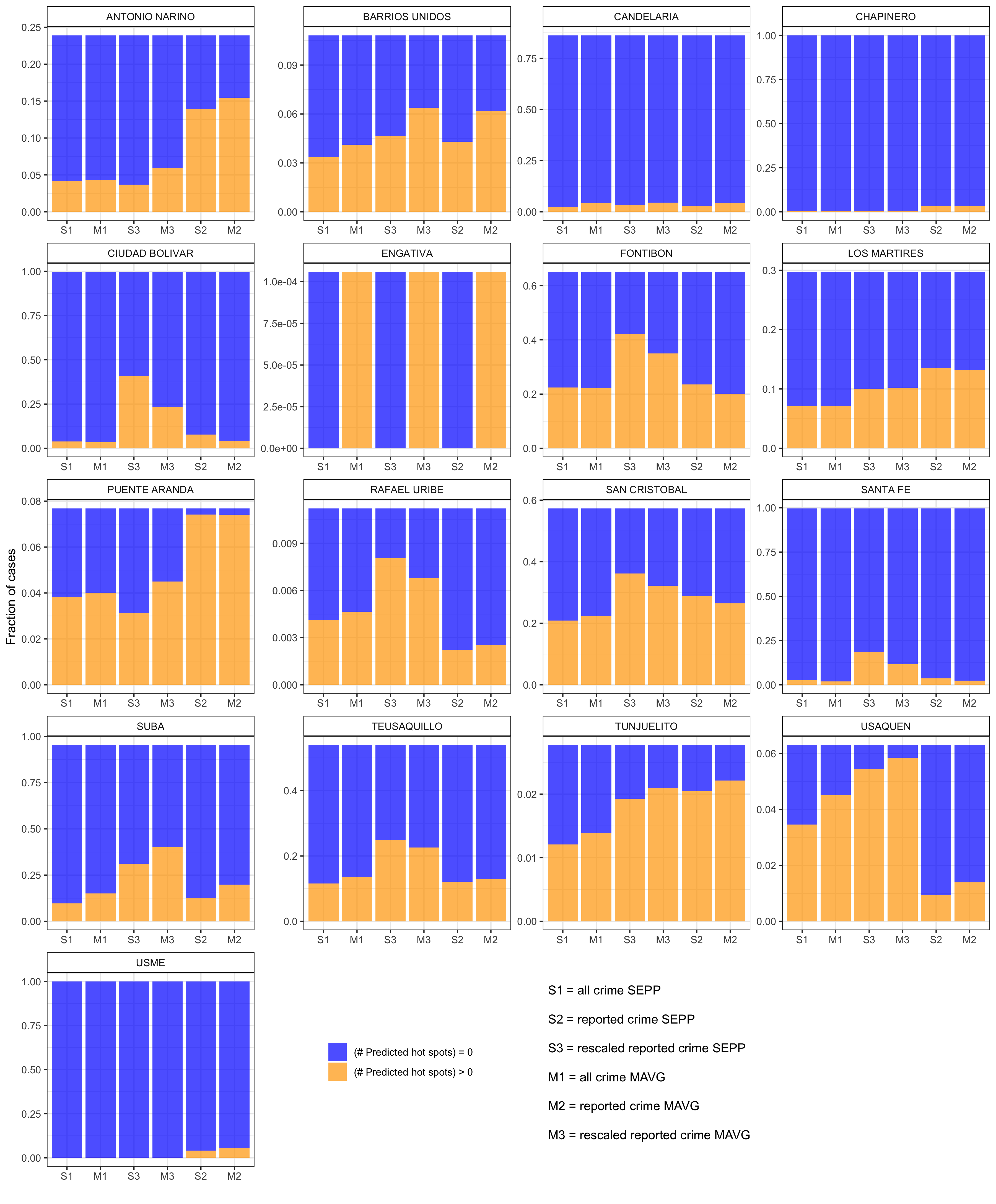}
    \caption{Fraction of prediction time steps with no true hot spots in district separated into instances with predicted and no predicted hot spots. Ratios are computed over all evaluation time steps (189 days) and all simulation runs (50 runs) with 50 hot spots selected at each step.}
    \label{fig:rel_count_no_true_appendix}
\end{figure*}

\begin{figure*}
    \centering
    \includegraphics[scale = 0.13]{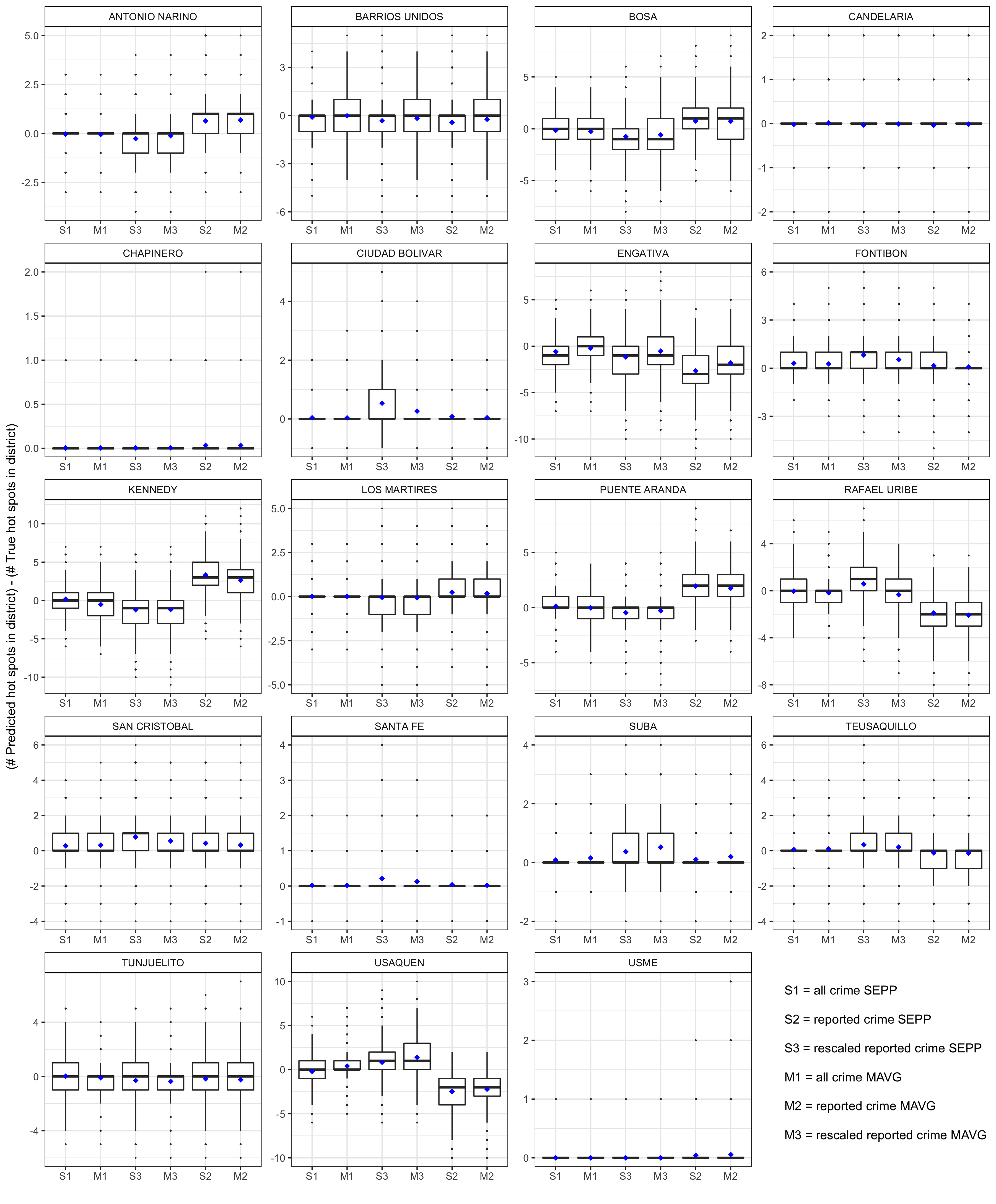}
    \caption{Absolute number of overpredicted hot spots over all evaluation time steps (189 days) and all simulation runs (50 runs) with 50 hot spots selected at each step.}
    \label{fig:diff_count_appendix}
\end{figure*}

\begin{figure*}
    \centering
    \includegraphics[scale = 0.13]{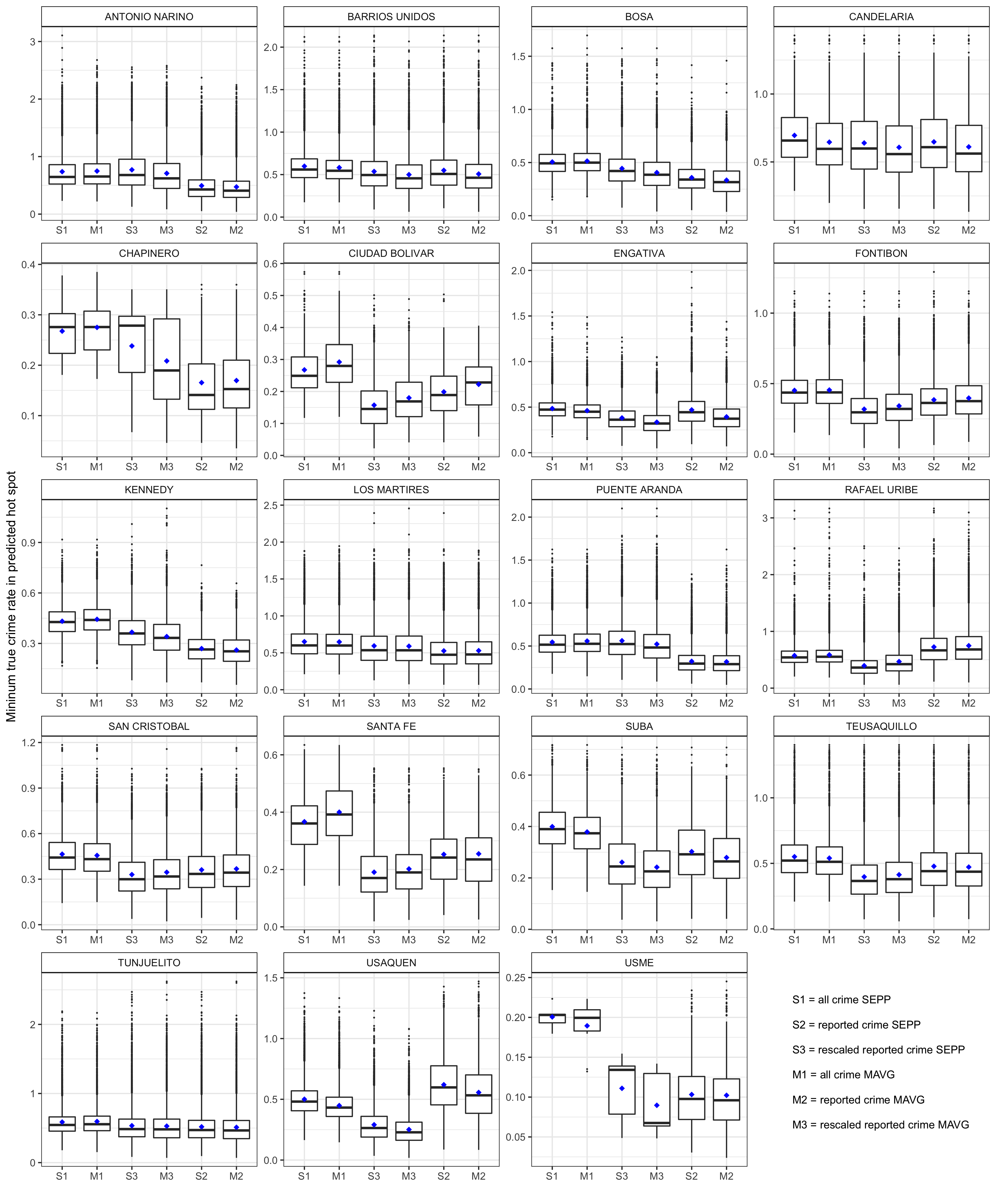}
    \caption{True crime thresholds for hot spot selection in Bogot\'a districts.
    Each point corresponds to an evaluation time step (189 days) and a simulation run (50 runs).
    A total of 50 hot spots is selected at each step, and cases in which no hot spot is predicted within the district are omitted for visualization.}
    \label{fig:rel_min_cg_appendix}
\end{figure*}


\end{document}